\documentstyle[aps,12pt]{revtex}

\renewcommand{\thefootnote}{\alph{footnote}}

\begin{document}

\renewcommand{\thefootnote}{\alph{footnote}}

\preprint{EFUAZ FT-97-53-REV}

\title{PHOTON-NOTOPH EQUATIONS\thanks{Contributed
to ``International Workshop on Fundamental Open Problems in
Mathematics, Physics and Other Sciences at the Turn of the Millenium",
Beijing, China, August 28, 1997 (Corresponding Participant).}}

\author{{\bf Valeri V. Dvoeglazov}}

\address {Escuela de F\'{\i}sica, Universidad Aut\'onoma de Zacatecas \\
Apartado Postal C-580, Zacatecas 98068, ZAC., M\'exico\\
Internet address:  valeri@cantera.reduaz.mx\\
URL: http://cantera.reduaz.mx/\~~valeri/valeri.htm}

%\date{Revised version: March 21, 1998}

\date{Received  \qquad\qquad 1998}

\maketitle

\medskip

\begin{abstract}

\baselineskip14pt

In the sixties Ogievetskii and Polubarinov proposed
the concept of a {\it notoph}, whose helicity
properties are complementary to those of a {\it photon}.
We analyze the theory of antisymmetric tensor fields
in the view of the normalization problem. The obtained
result is that it is possible to describe both photon and
notoph degrees of freedom on the basis of
the modified Bargmann-Wigner formalism for the symmetric
second-rank spinor. Physical consequences are discussed.
\end{abstract}

\newpage

\baselineskip15pt

In a recent series of the papers~\cite{DVO,DVO961,DVO962,DVO970,DVO97},
which are the continuation of the Ahluwalia {\it et al}
work~\cite{Ahl,DVA,DVALF} we tried to construct a self-consistent theory
of the quantized antisymmetric tensor (AST) field of the second rank and
of the 4-vector field. Previous published works
~\cite{DEB,Og,Hayh,Kalb,Ohanian,Avdeev}, as well as
textbooks~\cite{Lurie,Bogol,Novozh,Itzyk} can {\it not} be considered as
the works which solved the main problems, whether the quantized AST field
and the quantized 4-vector field are transverse or longitudinal fields
(in the sense if the helicity $h=\pm 1$ or $h=0$)?
can the electromagnetic potential be a 4-vector in a quantized theory
(cf.~\cite[p.251]{Wein})? how should the massless limit be taken?
and many other fundamental problems. The most rigorous
works are refs.~\cite{BW,WEIN0,Kim,Wein}, but it is not
easy to extract corresponding answers even from them. A lot of
problems of rigorous description of the light is still
opened. Ideas of this paper are based on three
referee reports from ``Foundation of Physics", which were very useful
even though critical ones.

First of all, we note after the referee that 1) ``...In natural units
($c=\hbar=1$) ... a lagrangian density, since the action is dimensionless,
has dimension of [energy]$^4$"; 2) One can always renormalize the
lagrangian density and ``one can obtain the same equations of motion...
by substituting $L \rightarrow (1/M^N) L$, where $M$ is an arbitrary
energy scale",  cf.~\cite{DVO962}; 3) the right physical dimension of the
field strength tensor $F^{\mu\nu}$ is [energy]$^2$; ``the transformation
$F^{\mu\nu} \rightarrow (1/2m) F^{\mu\nu}$ [which was regarded in
ref.~\cite{DVO97}] ... requires a more detailed study ... [because] the
transformation above changes its physical dimension: it is not a simple
normalization transformation". Furthermore, in the first papers on the
notoph~\cite{Og,Hayh,Kalb}\footnote{It is also known as a longitudinal
Kalb-Ramond field, but the consideration of Ogievetskii and Polubarinov
seems to me to be more rigorous
because it permits to study the $m\rightarrow 0$ procedure.} the authors
used the normalization of the 4-vector $F^\mu$ field\footnote{It is well
known that it is related to a third-rank antisymmetric field tensor.} to
[energy]$^2$ and, hence, the antisymmetric tensor ``potentials"
$A^{\mu\nu}$, to [energy]$^1$.

After taking into account these observations let us repeat the procedure
of the derivations of the Proca equations from the Bargmann-Wigner
equations for a {\it symmetric} second-rank spinor. We set
\begin{equation}
\Psi_{\{\alpha\beta\}} = (\gamma^\mu R)_{\alpha\beta} (c_a m A_\mu +
c_f  F_\mu) +(\sigma^{\mu\nu} R)_{\alpha\beta} (c_A m\gamma^5
A_{\mu\nu} + c_F F_{\mu\nu})\, ,\label{si}
\end{equation}
where
\begin{equation}
R=\pmatrix{i\Theta & 0\cr 0&-i\Theta\cr}\quad,\quad \Theta = -i\sigma_2 =
\pmatrix{0&-1\cr 1&0\cr}\, .
\end{equation}
Matrices $\gamma^{\mu}$ are chosen in the Weyl representation, i.e.,
$\gamma^5$ is assumed to be diagonal.  Constants $c_i$ are some numerical
dimensionless coefficients. The reflection operator $R$ has the properties
\begin{mathletters}
\begin{eqnarray}
&& R^T = -R\,,\quad R^\dagger =R =
R^{-1}\,,\\ && R^{-1} \gamma^5 R = (\gamma^5)^T\,,\\ && R^{-1}
\gamma^\mu R = -(\gamma^\mu)^T\,,\\ && R^{-1} \sigma^{\mu\nu} R = -
(\sigma^{\mu\nu})^T\,.
\end{eqnarray}
\end{mathletters}
They are  necessary for the expansion (\ref{si}) to be possible in such a
form, i.e., in order the $\gamma^{\mu} R$, $\sigma^{\mu\nu} R$
and $\gamma^5 \sigma^{\mu\nu} R$ to be {\it symmetrical} matrices.

The substitution of the above expansion into the Bargmann-Wigner
set~\cite{Lurie}
\begin{mathletters}
\begin{eqnarray}
\left [ i\gamma^\mu
\partial_\mu -m \right ]_{\alpha\beta} \Psi_{\{\beta\gamma\}} (x) &=&
0\,,\label{bw1}\\
\left [ i\gamma^\mu \partial_\mu -m \right
]_{\gamma\beta} \Psi_{\{\alpha\beta\}} (x) &=& 0\, . \label{bw2}
\end{eqnarray}
\end{mathletters}
gives us the new ``Proca" equations:
\begin{mathletters}
\begin{eqnarray}
&& c_a m (\partial_\mu A_\nu - \partial_\nu A_\mu ) +
c_f (\partial_\mu F_\nu -\partial_\nu F_\mu ) =
ic_A m^2 \epsilon_{\alpha\beta\mu\nu} A^{\alpha\beta} +
2 m c_F F_{\mu\nu} \, \label{pr1} \\
&& c_a m^2 A_\mu + c_f m F_\mu =
i c_A m \epsilon_{\mu\nu\alpha\beta} \partial^\nu A^{\alpha\beta} +
2 c_F \partial^\nu F_{\mu\nu}\, . \label{pr2}
\end{eqnarray}
\end{mathletters}
In the case $c_a=1$, $c_F ={1\over 2}$ and $c_f=c_A=0$ they
are reduced to the ordinary Proca equations.\footnote{We still note
that the division by $m$ in the first equation
is {\it not} a  well-defined operation in the case if someone
is interested in the subsequent limiting procedure $m\rightarrow 0$.
Probably, in order to avoid this obscure point one may wish
to write the Dirac equations in the form $\left [ (i\gamma^\mu
\partial_\mu)/m - \openone \right ] \psi (x) =0$ which
follows straightforwardly in the derivation of the Dirac equation
on the basis of the Ryder-Burgard relation~\cite{DVA}  and the Wigner
rules for boosting the field function from the zero-momentum frame.}
In the general case we obtain
dynamical equations which connect the photon, the notoph and
their potentials. Divergent (in $m\rightarrow 0$) parts
of field functions and of dynamical variables
should be removed by corresponding gauge (or Kalb-Ramond gauge)
transformations. It is well known that the notoph massless field is
considered to be the pure
longitudinal field after one takes into account $\partial_\mu A^{\mu\nu}=
0$.  Apart from these dynamical equations we can obtain the set of
constraints by means of the subtraction of the equations of the
Bargmann-Wigner set (instead of the addition as for
(\ref{pr1},\ref{pr2})). It reads \begin{mathletters} \begin{eqnarray}
&&mc_a \partial^\mu A_\mu + c_f \partial^\mu f_\mu =0\, , \\
&&mc_A \partial^\alpha A_{\alpha\mu} + {i\over 2}
c_F \epsilon_{\alpha\beta\nu\mu}
\partial^\alpha F^{\beta\nu} = 0\, .
\end{eqnarray}
\end{mathletters}
that suggests $\widetilde F^{\mu\nu} \sim im A^{\mu\nu}$
and $f^\mu \sim mA^\mu$, as in~\cite{Og}.

Thus, after the suitable choice of the dimensionless coefficients
$c_i$ the
lagrangian density for the photon-notoph field can be proposed:
\begin{eqnarray}
{\cal L} &=& {\cal L}^{Proca} +{\cal L}^{Notoph} = -
{1\over 8} F_\mu F^\mu -{1\over 4} F_{\mu\nu} F^{\mu\nu} +\nonumber\\
&+& {m^2 \over 2} A_\mu A^\mu + {m^2 \over 4} A_{\mu\nu} A^{\mu\nu}\, , \
\end{eqnarray}
The limit $m\rightarrow 0$ may be taken for dynamical variables,
in the end of calculations only.

Furthermore, it is logical to introduce the normalization scalar field
$\varphi (x)$ and consider the expansion:
\begin{equation}
\Psi_{\{\alpha\beta\}} = (\gamma^\mu R)_{\alpha\beta} (\varphi A_\mu)
+ (\sigma^{\mu\nu} R)_{\alpha\beta} F_{\mu\nu}\, .
\end{equation}
Then, we arrive at the following set
\begin{mathletters}
\begin{eqnarray}
&&2m F_{\mu\nu} = \varphi (\partial_\mu A_\nu -\partial_\nu A_\mu)
+ (\partial_\mu \varphi) A_\nu - (\partial_\nu \varphi) A_\mu\, ,\\
&& \partial^\nu F_{\mu\nu} = {m \over 2} (\varphi A_\mu)\, ,
\end{eqnarray}
\end{mathletters}
which in the case of the constant
scalar field $\varphi = 2m$ also can be reduced to the set of the
Proca equations. The additional constraints are
\begin{mathletters}
\begin{eqnarray}
&&(\partial^\mu \varphi) A_\mu + \varphi (\partial^\mu A_\mu) =0\,,\\
&& \partial_\mu \widetilde F^{\mu\nu} = 0\, .
\end{eqnarray}
\end{mathletters}

At the moment it is not yet obvious how can we
account for other equations in the $(1,0)\oplus (0,1)$
representation, e.g.~[7b].   One can wish to seek the generalization of
the Proca set on the basis of the introduction of two mass parameters
$m_1$ and $m_2$. But, when we apply the BW procedure to the Dirac
equation we cannot obtain new physical content.  Another equation in the
$(1/2,0)\oplus (0,1/2)$ representation was obtained in
ref.~\cite{Raspini}.  It has the form:
\begin{equation} \left [
i\gamma^\mu \partial_\mu - m_1 - \gamma^5 m_2 \right ] \Psi (x) =0\,.
\end{equation}
The Bargmann-Wigner procedure for the set of this kind of
equations (which include the $\gamma^5$ matrix in the mass term) yields:
\begin{mathletters}
\begin{eqnarray}
&&2m_1  F^{\mu\nu} +2i m_2 \widetilde F^{\mu\nu} = \varphi
(\partial^\mu A^\nu -\partial^\nu A^\mu) +
(\partial^\mu \varphi) A^\nu - (\partial^\nu \varphi) A^\mu\, ,\\
&&\partial^\nu F_{\mu\nu} = {m_1 \over 2} (\varphi A_\mu) \,
\end{eqnarray}
\end{mathletters}
with the constraints
\begin{mathletters}
\begin{eqnarray}
&&(\partial^\mu \varphi) A_\mu +\varphi (\partial^\mu A_\mu) = 0\, \\
&&\partial^\nu \widetilde F_{\mu\nu} = {im_2 \over 2} (\varphi A_\mu)\, .
\end{eqnarray}
\end{mathletters}
The equality of mass factors ($m_1^{(1)} = m_1^{(2)}$
and $m_2^{(1)} = m_2^{(2)}$) in the set of the
Dirac equations is obtained in the process of calculations
as necessary conditions.

In fact, the results of this paper develop the
old results of ref.~\cite{Og}.  We returned to this question due to recent
interpretational controversies in claims of experimental observations of
the objects ${\bf E} \times {\bf E}^\ast$ and ${\bf A} \times {\bf
A}^\ast$ in the non-linear optics~\cite{exper}.\footnote{One can wish to
compare the {\it notoph} concept with theoretical works of M. W. Evans
{\it et al.} [e.g., {\it The Enigmatic Photon}, Vols. I-IV, Kluwer Academic
Publishers, 1994-97]. It is easy to see from the formulas (9,10) of
ref.~\cite{Og} that the Evans' proposal is {\it not any novelty}. The
longitudinal field constructed from {\it polarization vectors} is nothing
more than the {\it notoph} antisymmetric tensor
potentials~\cite[Eq.(10)]{Og}. On the other hand, in the Evans' ${\bf B}$
cyclic relations we found that ${\bf B}^{(3)}$ field is {\it not} a part
of the {\it antisymmetric} tensor due to different Lorentz
transformations~\cite{DVO970}. While Evans refers often to ${\bf A}\times
{\bf A}^\ast$ and ${\bf E}\times {\bf E}^\ast$ (or $\sim {\bf
B}^{(1)}\times {\bf B}^{(2)}$) as the same entities in {\it all cases},
this is not so.
These Evans' claims are contradictory each other in the view of
the Lorentz symmetry.  While, in my opinion,
Evans' works on the theory of longitudinal modes of
electromagnetism are full of errors,
the old work of V. I. Ogievetskii and I. V.  Polubarinov
shows that one can work rigorously with  these concepts.} In this
connection one can consider that $\sim {\bf A}\times {\bf A}^\ast$ term
can be regarded as the part of antisymmetric tensor potential and $\sim
{\bf B}\times {\bf B}^\ast$, as the part of the 4-vector field (cf.  the
formulas (19a-c) in ref.~\cite{DVO97}). According to~\cite[Eqs.(9,10)]{Og}
we proceed in the construction of the ``potentials" for the notoph as
follows:
\begin{equation} A_{\mu\nu} ({\bf p})  = N \left
[\epsilon_\mu^{(1)} ({\bf p})\epsilon_\nu^{(2)} ({\bf p})-
\epsilon_\nu^{(1)} ({\bf p}) \epsilon_\mu^{(2)} ({\bf p}) \right ]
\end{equation} On using explicit forms for the polarization vectors in the
momentum space (e.g., refs.~\cite{Wein} and~\cite[formulas(15a,b)]{DVO97})
one obtains
\begin{eqnarray} A^{\mu\nu} ({\bf p}) = {iN^2 \over m} \pmatrix{0&-p_2&
p_1& 0\cr p_2 &0& m+{p_r p_l\over p_0+m} & {p_2 p_3\over p_0 +m}\cr -p_1 &
-m - {p_r p_l \over p_0+m}& 0& -{p_1 p_3\over p_0 +m}\cr 0& -{p_2 p_3
\over p_0 +m} & {p_1 p_3 \over p_0+m}&0\cr}\, , \label{lc}
\end{eqnarray}
i.e., it coincides with the longitudinal components of the antisymmetric
tensor obtained in refs.~[7a,Eqs.(2.14,2.17)]
and~\cite[Eqs.(17b,18b)]{DVO97}  within the normalization and
different forms of the spin basis.  The longitudinal
states reduce to zero in the
massless case under appropriate choice of the normalization and only if a
$j=1$ particle moves along with the third axis $OZ$. It is also
useful to compare Eq. (\ref{lc}) with the formula (B2) in ref.~\cite{DVALF}
in order to realize the correct procedure for taking the massless limit.

Next, the Tam-Happer experiments~\cite{TH} did not
find satisfactory explanation in the framework of the ordinary QED (at
least, their explanation is complicated by huge technical calculations).
On the other hand, in ref.~\cite{Pradhan} the very interesting model has
been proposed.  It is based on
gauging the Dirac field on using the coordinate-dependent parameters
$\alpha_{\mu\nu} (x)$ in
\begin{equation}
\psi(x) \rightarrow \psi^\prime (x^\prime) = \Omega \psi(x)\,\,, \quad
\Omega = \exp \left [ {i\over 2} \sigma^{\mu\nu} \alpha_{\mu\nu}(x)
\right ]\, .
\end{equation}
and, thus,  the second ``photon" was introduced. The
compensating 24-component (in general) field $B_{\mu,\nu\lambda}$ reduces
to the 4-vector field as follows (the notation of~\cite{Pradhan} is used
here):
\begin{equation}
B_{\mu,\nu\lambda} = {1\over 4} \epsilon_{\mu\nu\lambda\sigma} a_\sigma
(x) \, .
\end{equation}
As readily seen after the comparison of these formulas with those of
refs.~\cite{Og,Hayh,Kalb}, the second photon is nothing more than the
Ogievetskii-Polubarinov {\it notoph} within the normalization.
Parity properties (as well as its behavior in the
massless limit) are dependent not only on the explicit forms of the
momentum-space field functions of the $(1/2,1/2)$ representation, {\it
but} also on the properties of corresponding creation/annihilation
operators. Helicity properties depend on the normalization.

Finally, in my opinion, the recent theoretical concepts of {\it
action-at-a-distance} reposed by A. E. Chubykalo {\it et al.}, e.g.,
ref.~\cite{Chub} and the concept of flavour-oscillation clocks governed
by scalar gravitational potential~\cite{grav} should find connections
with the longitudinal quantum fields.

\acknowledgements
This paper has been inspired by remarks of the
referees of IJMPA (1994) and {\it Foundation of Physics} (1998),
by the papers of Prof.  D.  V.  Ahluwalia and his predecessors
who worked with the theories of antisymmetric tensor fields.
I am obliged to Profs. L. V. Avdeev, A. E. Chubykalo,
S. Esposito, Y. S.  Kim,
A.  F.  Pashkov, S. Roy and Yu. F. Smirnov for
illuminating discussions.

I am grateful to Prof. R. Santilli  for his kind invitation to write the
paper for the present Workshop.

I am grateful to Zacatecas University for awarding a professorship.
This work has been supported in part by the Mexican Sistema
Nacional de Investigadores and by the CONACyT, M\'exico under the research
project~0270P-E.

\end{document}